\documentclass[a4paper,11pt]{article}
\usepackage{pos}
\usepackage{bm}
\usepackage{braket}
\usepackage{amsmath}
\usepackage{graphicx}% Include figure files

\title{Most charming dibaryon near unitarity}
\author*[a,b]{Yan Lyu}
\author[a,c]{Hui Tong}
\author[c]{Takuya Sugiura}
\author[d, b]{Sinya Aoki}
\author[b,c]{Takumi Doi}
\author[c]{Tetsuo Hatsuda}
\author[a,e]{Jie Meng}
\author[b]{Takaya Miyamoto}

\affiliation[a]{State Key Laboratory of Nuclear Physics and Technology, School of Physics, Peking University, Beijing 100871, China }
\affiliation[b]{Quantum Hadron Physics Laboratory, RIKEN Nishina Center, Wako 351-0198, Japan}
\affiliation[c]{Interdisciplinary Theoretical and Mathematical Sciences Program (iTHEMS), RIKEN, Wako 351-0198, Japan}
\affiliation[d]{Center for Gravitational Physics, Yukawa Institute for Theoretical Physics, Kyoto University, Kyoto 606-8502, Japan}
\affiliation[e]{Yukawa Institute for Theoretical Physics, Kyoto University, Kyoto 606-8502, Japan}

\emailAdd{helvetia@pku.edu.cn}
\emailAdd{tong16@pku.edu.cn}
\emailAdd{takuya.sugiura@riken.jp}
\emailAdd{saoki@yukawa.kyoto-u.ac.jp}
\emailAdd{doi@ribf.riken.jp}
\emailAdd{thatsuda@riken.jp}
\emailAdd{mengj@pku.edu.cn}
\emailAdd{miyamoto@ribf.riken.jp}

\abstract{We present a first study on a pair of triply charmed baryons, $\Omega_{ccc}\Omega_{ccc}$ in the $^1S_0$ channel, on the basis of the 
HAL QCD method. 
The measurements are perfomed on the $(2+1)$-flavor lattice QCD configurations with nearly physical light-quark masses and physical charm-quark mass. 
We show that the system with the Coulomb repulsion taking into account the charge form factor of $\Omega_{ccc}$ leads to the scattering length $a^\mathrm{C}_0\simeq-19$ fm and the effective range $r^\mathrm{C}_\mathrm{eff}\simeq0.45$ fm, 
which indicates $\Omega_{ccc}\Omega_{ccc}$ is located in the unitary regime.}

\FullConference{%
 The 38th International Symposium on Lattice Field Theory, LATTICE2021
  26th-30th July, 2021
  Zoom/Gather@Massachusetts Institute of Technology
}

\begin{document}
\raisebox{80pt}[0pt][0pt]{\hspace*{50mm} RIKEN-QHP-512, RIKEN-iTHEMS-Report-21, YITP-21-149}
\vspace{-12mm}
%\begin{flushright}
%  RIKEN-QHP-512, RIEKN-iTHEMS-Report-21, YITP-21-149 
%\end{flushright}
\maketitle

\section{Introduction}
Quest for dibaryon is one of the most challenging problems in nuclear and particle physics~\cite{Clement2017, Gal2015}.
Experimentally, the deuteron, composed of a proton and a neutron, is the only stable bound state. 
There are possible dibaryons composed of light quarks ($u$, $d$ and $s$).
In particular, a theoretical progress on the basis of lattice QCD (LQCD) simulations near the physical point has been made recently on the dibaryons with many strangeness,
$p\Omega$($uudsss$)~\cite{Iritani2019} and $\Omega\Omega$($ssssss$)~\cite{Gongyo2018}.

Going beyond the light-quark sector, there are several studies on the charmed dibaryons from both phenomenological approach and LQCD approach~\cite{Huang2020, Richard2020, Junnarkar2019}.
In the charm-quark sector, although so far only the singly charmed baryon~\cite{Aaij2017Omega} and the doubly charmed baryon~\cite{Aaij2017cascade} have been observed,
the triply charmed baryon $\Omega_{ccc}$ provides an ideal system to study the perturbative and nonperturbative aspects of QCD in the baryonic sector.
Since it is predicted by Bjorken in 1980s~\cite{Bjorken1985}, there are many studies on its mass and electromagnetic form factor (see~\cite{Can2015} and references therein).

In this paper, we explore the charmed dibaryons, by focusing on its simplest possible form, $\Omega_{ccc}\Omega_{ccc}$ in the $^1S_0$ channel, from a first principle LQCD approach.
In this specific channel, the maximum attraction is expected due to the Pauli exclusion between charm quarks at short distances does not operate when spin $s=0$ and angular momentum $L=0$.
The HAL QCD method~\cite{Aoki2020, Ishii2007, Ishii2012} is used to convert the spatial baryon-baryon correlation to the scattering parameters.
As we will show latter, the $\Omega_{ccc}\Omega_{ccc}$ in the $^1S_0$ channel with the Coulomb repulsion is located near unitarity~\cite{Lyu2021}.

\section{HAL QCD method}
The equal-time Nambu-Bethe-Salpeter (NBS) amplitude $\psi(\bm r)$, whose asymptotic behavior at large distances reproduces the scattering phase shift,
plays an important role in the HAL QCD method~\cite{Aoki2020, Ishii2007, Ishii2012}, from which a non-local but energy-independent potential $U(\bm r,\bm r')$ can be defined. 
Since all the elastic scattering states are governed by the same potential $U(\bm r,\bm r')$,
the time-dependent HAL QCD method~\cite{Ishii2012} takes full advantage of all the NBS amplitudes below the inelastic threshold $\Delta E^*\sim\Lambda_\mathrm{QCD}$ by
defining the $R$ correlator as follows.
\begin{equation}
 \begin{split}
  R(\bm r, t>0)&=\sum_{\bm x}\braket{0|\hat\Omega_{ccc}(\bm x, t)\hat\Omega_{ccc}(\bm r+\bm x, t)\overline{\mathcal J}(0)|0}/e^{-2m_{\Omega_{ccc}}t}\\
 &=\sum_n A_n\psi_n(\bm r)e^{-(\Delta W_n)t} +O(e^{-(\Delta E^*)t}),
 \end{split}
\end{equation}
with local interpolating operators $\hat\Omega_{ccc}$ and a wall-type source operator $\overline{\mathcal J}$.
$A_n$ is the overlapping factor defined by $\braket{n|\overline{\mathcal J}(0)|0}$, with $\ket{n}$ representing the QCD eigensates in a finite volume below the inelastic threshold,
and $\Delta W_n =2\sqrt{m_{\Omega_{ccc}}^2 + \bm k^2_n}-2m_{\Omega_{ccc}}$ with the baryon mass $m_{\Omega_{ccc}}$ and the relative momentum $\bm k_n$.
The contributions from the inelastic states are exponentially suppressed when $t\gg(\Delta E^*)^{-1}$. 
As long as the condition for $t$ is met, the $R$ correlator can be shown to satisfy following integro-differential equation~\cite{Ishii2012},
\begin{equation}
 \left(\frac{1}{4m_{\Omega_{ccc}}}\frac{\partial^2}{\partial t^2}-\frac{\partial}{\partial t}+\frac{\nabla^2}{m_{\Omega_{ccc}}}\right)R(\bm r,t)=\int d\bm r' U(\bm r,\bm r')R(\bm r', t).
\end{equation}
The effective central potential in the leading order approximation of the derivative expansion,
$U(\bm r,\bm r')=V(r)\delta(\bm r-\bm r')+\sum_{n=1}V_{2n}(\bm r)\nabla^{2n}(\bm r-\bm r')$, is given by,
\begin{equation}
 V(r)=\frac{1}{R(\bm r, t)}\left(\frac{1}{4m_{\Omega_{ccc}}}\frac{\partial^2}{\partial t^2}-\frac{\partial}{\partial t}+\frac{\nabla^2}{m_{\Omega_{ccc}}}\right)R(\bm r,t).
\end{equation}

\section{Lattice setup}
$(2+1)$-flavor gauge configurations are generated on the $96^4$ lattice with the Iwasaki gauge action at $\beta=1.82$
and nonperturbatively $O(a)$-improved Wilson quark action with $c_\mathrm{sw}=1.11$ and stout smearing~\cite{Ishikawa2016}. 
The lattice spacing is $a\simeq0.0846$ fm ($a^{-1}\simeq2.333$ GeV), the pion mass and the kaon mass are $m_\pi\simeq146$ MeV and $m_K\simeq525$ MeV, respectively.
We use the relativistic heavy quark (RHQ) action for the charm quark, which is designed to remove the leading order and the next-to-leading order cutoff errors~\cite{Aoki20013}.
Two sets of RHQ parameters~\cite{Namekawa2017} are used so as to interpolate the physical charm-quark mass and reproduce the dispersion relation for the spin-averaged $1S$ charmonium.
The interpolated mass for $\Omega_{ccc}$ is $m_{\Omega_{ccc}}\simeq4796$ MeV.

The measurements are performed by a combination of the Bridge++ code~\cite{Bridge} and the unified contraction algorithm~\cite{Doi2013},
where the former is used for the quark propagator and the latter is used for the contraction.
The periodic (Derichlet) boundary condition is employed for spatial (temporal) direction.
Four time measurements are performed by shifting the source position, where the Coulomb gauge fixing is imposed.
Forward and backward propagation are averaged to reduce the statistical fluctuations.
The total measurements is $112$ configurations $\times$ $4$ source positions $\times$ $2$ (forward and backward).

\section{Numerical results}

\begin{figure}[htb]
  \centering
  \includegraphics[width=9.5cm]{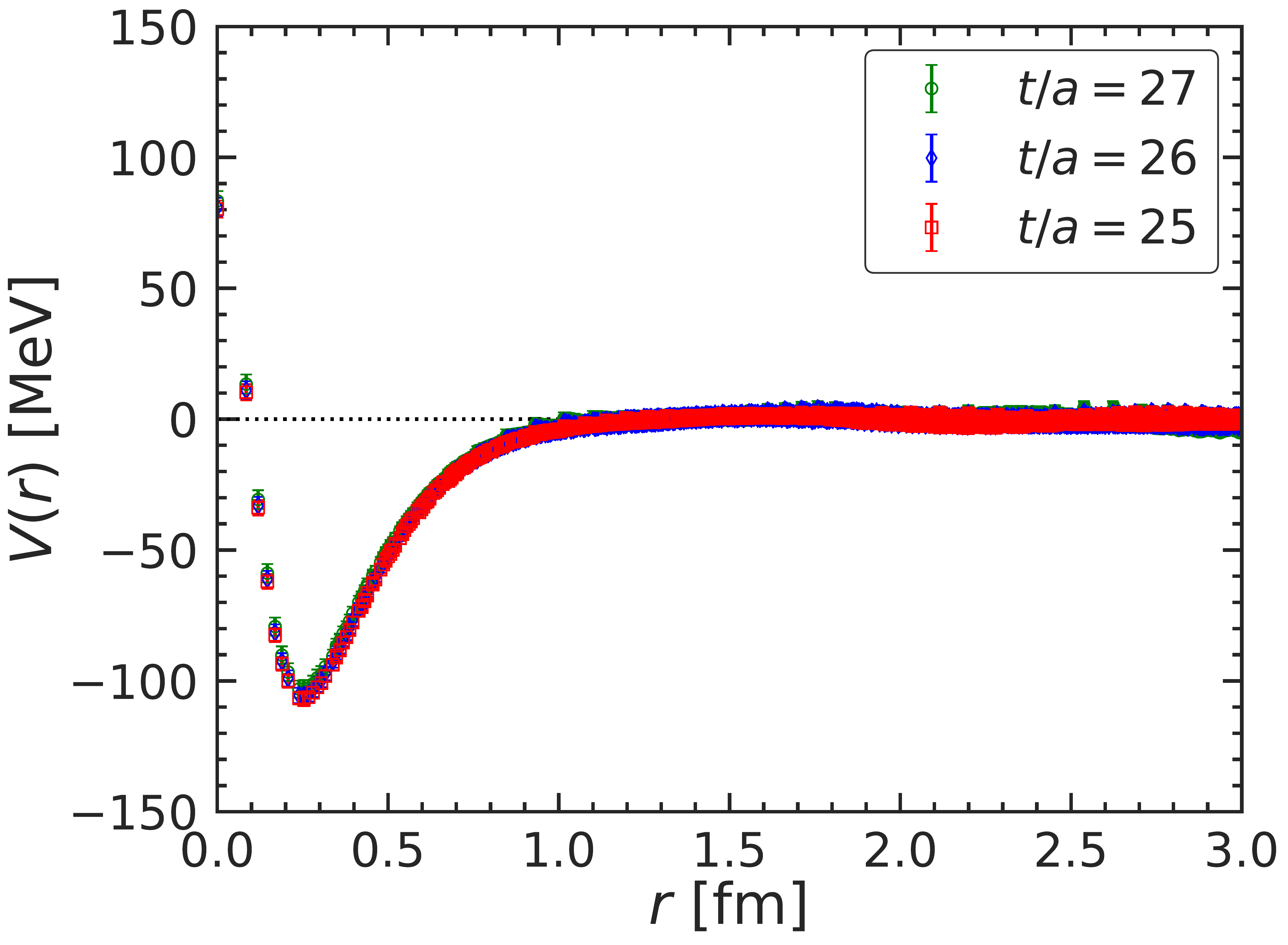}
  \caption{
  The $\Omega_{ccc}\Omega_{ccc}$ potential $V(r)$ in the ${^1S_0}$ channel at Euclidean time $t/a= 25$, $26$, and $27$.
  }
  \label{Fig1}
\end{figure}
In Fig.~\ref{Fig1}, we show the $\Omega_{ccc}\Omega_{ccc}$ potential $V(r)$ from the interpolation between two sets in the $^1S_0$ channel at Euclidean time $t/a=25$, $26$, and $27$.
The statistical errors for the potentials are estimated by the jackknife method with a bin size of 14 configurations.
The potentials for $t/a=25$, $26$, and $27$ are nearly identical within statistical errors, which indicates systematic errors due to truncation of the derivative expansion and the inelastic states are small~\cite{Ishii2012}.

The potential $V(r)$ has qualitative features similar to $NN$ potential~\cite{Doi2017} and $\Omega\Omega$ potential~\cite{Gongyo2018}, i.e., a repulsive core surrounded by an attractive well.
The magnitude of the repulsion for $\Omega_{ccc}\Omega_{ccc}$ is an order of magnitude smaller than that of $\Omega\Omega$,
which is in line with the inverse ratio of the square of their constituent quark mass from the perspective of color-magnetic interaction~\cite{Oka1987}.
In addition, the attraction, which may be attributed to the exchange of charmed mesons, is short-ranged.

\begin{figure}[htb]
  \centering
  \includegraphics[width=9.5cm]{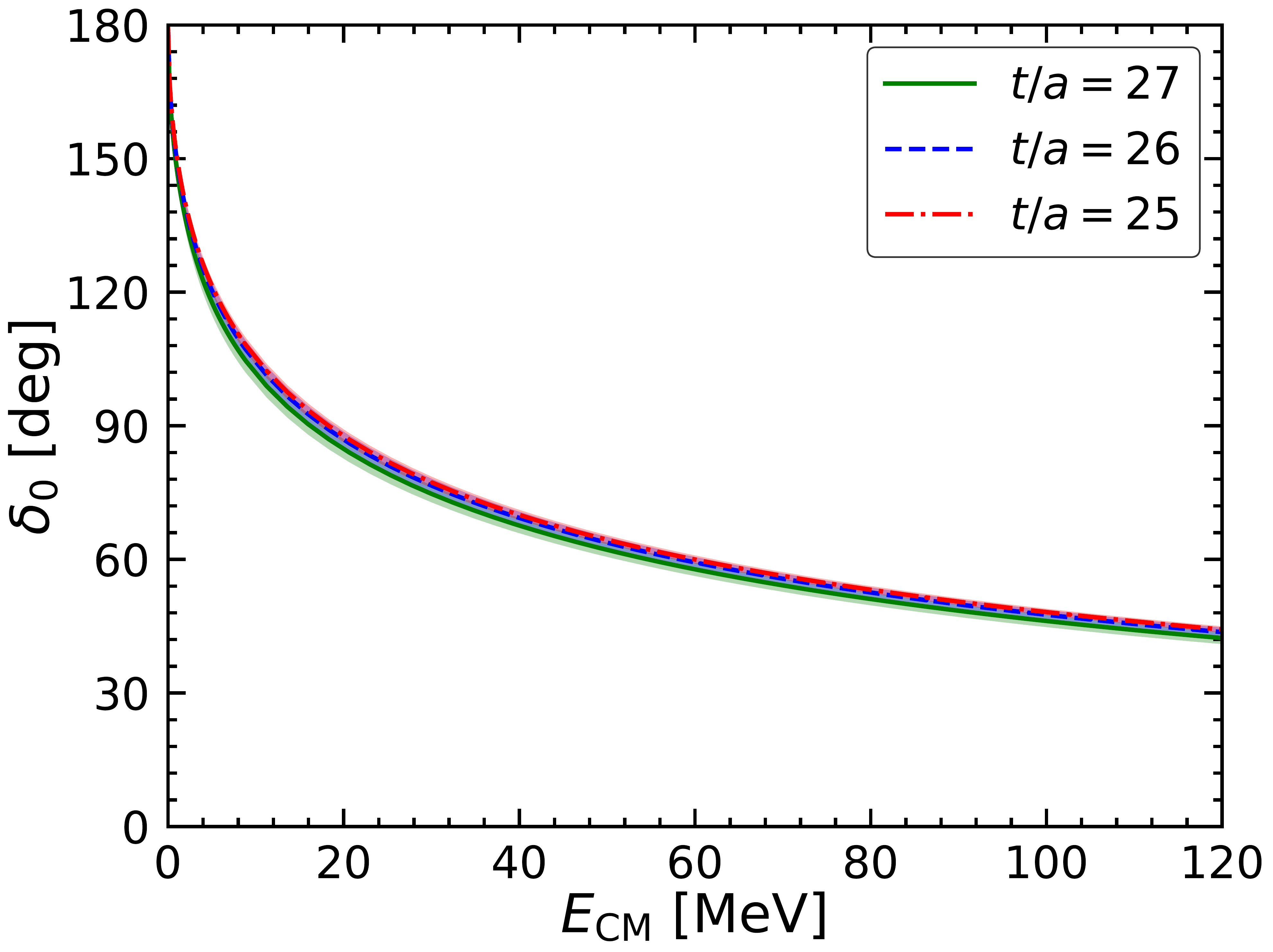}
  \caption{
  The scattering phase shifts $\delta_0$ in the ${^1S_0}$ channel for $t/a= 25$, $26$, and $27$ as a function of the kinetic energy $E_\mathrm{CM}=2\left(\sqrt{k^2+m_{\Omega_{ccc}}^2}-m_{\Omega_{ccc}}\right)$ in the center of mass frame.
  }
  \label{Fig2}
\end{figure}
The $\Omega_{ccc}\Omega_{ccc}$ scattering phase shifts $\delta_0$ in the $^1S_0$ channel calculated from the potential $V(r)$ at $t/a=25$, $26$, and $27$ are shown in Fig.~\ref{Fig2}.
The kinetic energy in the center of mass frame is defined as $E_\mathrm{CM}=2\left(\sqrt{k^2+m_{\Omega_{ccc}}^2}-m_{\Omega_{ccc}}\right)$.
The phase shifts start from $180^\circ$ indicates the existence of a bound state in the system without Coulomb repulsion.

Using the effective range expansion, $k\cot\delta_0=-1/a_0+1/2r_\mathrm{eff}k^2+O(k^4)$ with the scattering length $a_0$ and the effective range $r_\mathrm{eff}$, we extract the low energy parameters as follows.
\begin{equation}
  a_0=1.57(0.08)\left(^{+0.12}_{-0.04}\right)~\mathrm{fm}, \qquad r_\mathrm{eff}=0.57(0.02)\left(^{+0.01}_{-0.00}\right)~\mathrm{fm},
\end{equation}
where the central values and the statistical errors in the first parentheses are extracted from $\delta_0$ at $t/a=26$, while the systematics errors in the second parentheses are estimated from the results at $t/a=25$ and $27$.

\begin{figure}[htb]
  \centering
  \includegraphics[width=9.5cm]{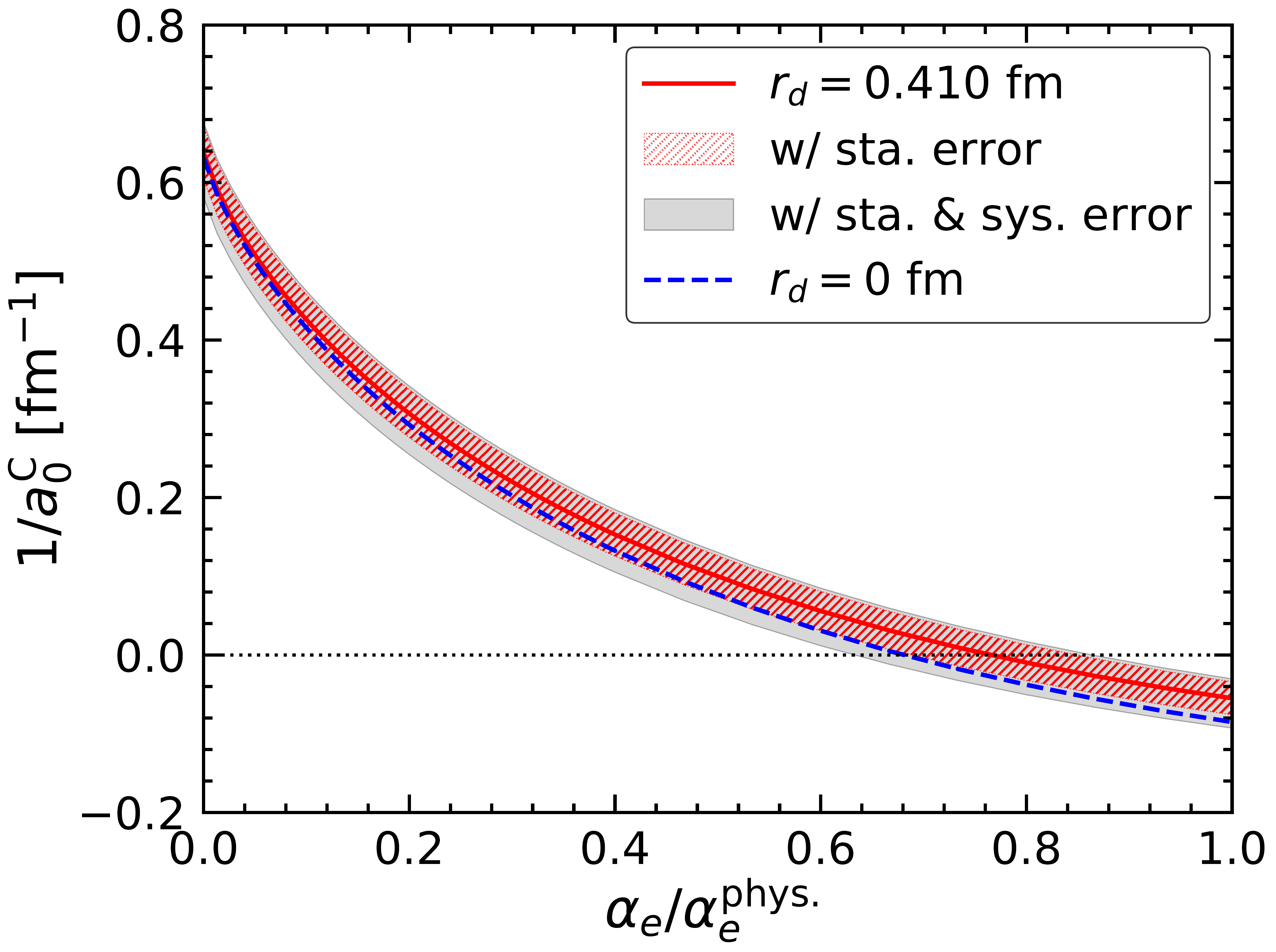}
  \caption{
  The inverse of the scattering length $1/a^\mathrm{C}_0$ as a function of $\alpha_e/\alpha_e^\mathrm{phys.}$.
  The red solid (blue dashed) line is the central values for $r_d=0.410$~fm~\cite{Can2015} ($r_d=0$~fm). 
  The inner (outer) band shows the statistical (statistical and systematic) errors.
  }
  \label{Fig3}
\end{figure}

The binding energy $B$ of the bound state is found to be $B\simeq5.7$ MeV, and the corresponding root-mean-square distance is $\sqrt{\braket{r^2}}\simeq1.1$~fm.
Since the binding energy and the size of the bound state from the strong interaction are not large, the Coulomb repulsion between two $\Omega_{ccc}^{++}$s needs to be taken into account.
Therefore, we consider the Coulomb repulsion $V^\mathrm{Coulomb}(r)$ by using the lattice results for the charge radius $r_d$ of $\Omega_{ccc}^{++}$ in Ref.~\cite{Can2015}, $V^\mathrm{Coulomb}(r)=4\alpha_e/rF(x)$ with $F(x)=1-e^{-x}(1+11x/16+3x^2/16+x^3/48)$ and $x=2\sqrt6r/r_d$.
In order to see how the Coulomb repulsion affects the system, we show the inverse of the scattering length $1/a^\mathrm{C}_0$ under the change of $\alpha_e$ from $0$ to its physical value $\alpha_e^\mathrm{phys.}=1/137.036$ in Fig.~\ref{Fig3}.
At the physical point, i.e., $\alpha_e/\alpha_e^\mathrm{phys.}=1$, the scattering parameters are,
\begin{equation}
  a^\mathrm{C}_0=-19(7)\left(^{+7}_{-6}\right)~\mathrm{fm},\qquad r^\mathrm{C}_\mathrm{eff}=0.45(0.01)\left(^{+0.01}_{-0.00}\right)~\mathrm{fm}.  
\end{equation}

\begin{figure}[htb]
  \centering
  \includegraphics[width=9.5cm]{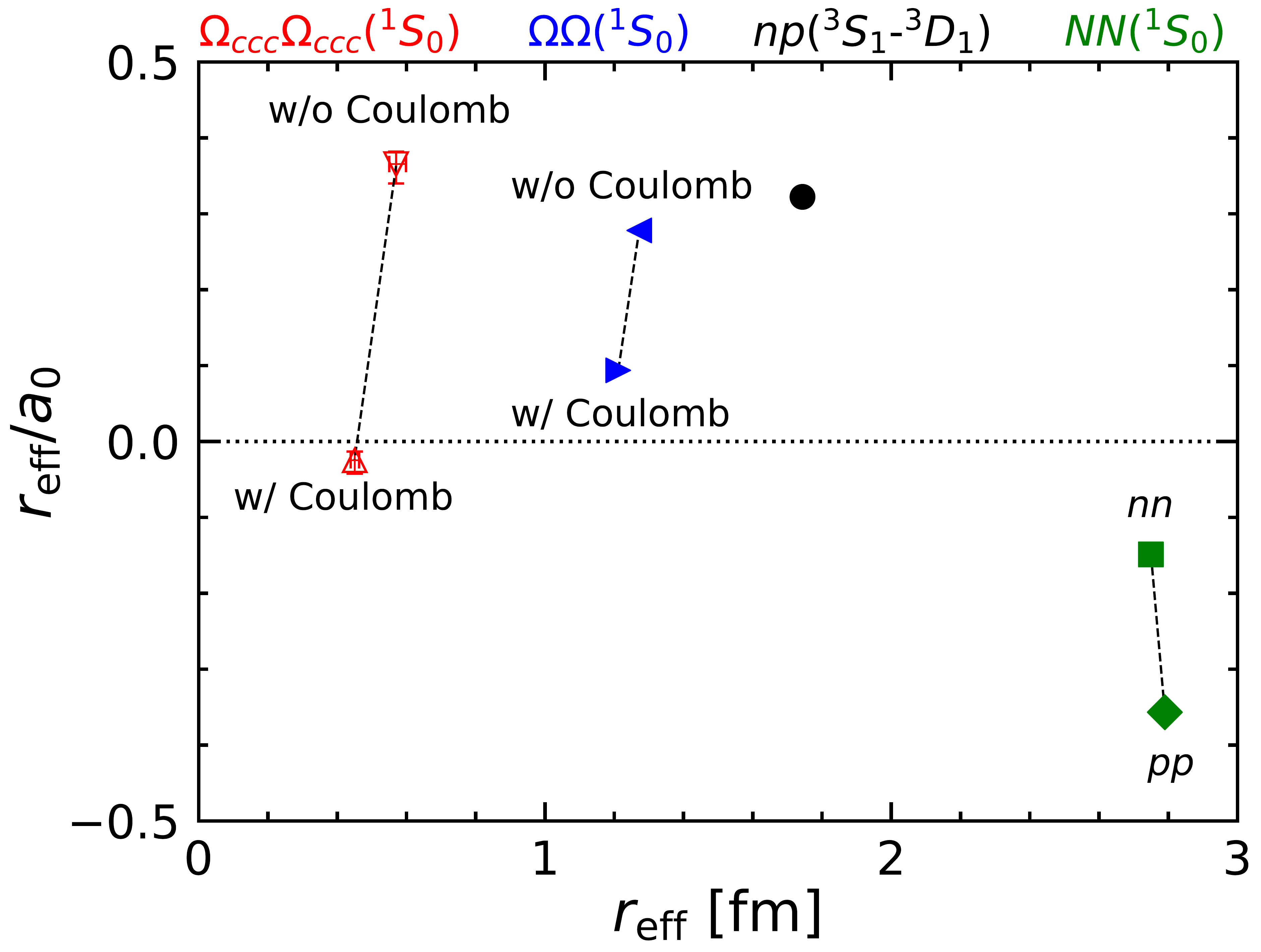}
  \caption{
    The dimensionless ratio $r_\mathrm{eff}/a_0$ as a function of $r_\mathrm{eff}$.
    The $\Omega_{ccc}\Omega_{ccc}$, $\Omega\Omega$, and $NN$ in the $^1S_0$ channel with (without) Coulomb repulsion
    are shown by the red up(down)-pointing triangle, blue right(left)-pointing triangle, and green diamond (square) respectively.
    The black circle represents the $NN(^3S_1-^3D_1)$.
  }
  \label{Fig4}
\end{figure}

We plot the dimensionless ratio $r_\mathrm{eff}/a_0$ as a function of $r_\mathrm{eff}$ in Fig.~\ref{Fig4}.
The $\Omega_{ccc}\Omega_{ccc}$, $\Omega\Omega$, and $NN$ in the $^1S_0$ channel with (without) Coulomb repulsion
are shown by the red up(down)-pointing triangle, blue right(left)-pointing triangle, and green diamond (square) respectively.
The black circle represents the $NN(^3S_1-^3D_1)$.
Among all those dibaryons, $\Omega^{++}_{ccc}\Omega^{++}_{ccc}(^1S_0)$ is the closest to unitarity.

\section{Summary}
The $\Omega_{ccc}\Omega_{ccc}$ in the $^1S_0$ channel is studied for the first time on the basis of $(2+1)$-flavor lattice QCD simulations with nearly physical
light-quark masses and physical charm-quark mass.
The potential from the time-dependent HAL QCD method leads to a most charming ($C=6$) dibaryon.
By taking into account the Coulomb repulsion, the $\Omega^{++}_{ccc}\Omega^{++}_{ccc}(^1S_0)$ is found to be located in the unitary regime.
It is an interesting future work to study $\Omega^-_{bbb}\Omega^-_{bbb}(^1S_0)$ to reveal the quark mass dependence of the scattering parameters.

\section*{Acknowledgements}
We thank the members of HAL QCD Collaboration, the members of PACS Collaboration, Yusuke Namekawa,
Tatsumi Aoyama, Haozhao Liang, Shuangquan Zhang and Pengwei Zhao for technical supports and stimulating  discussions.
We thank ILDG/JLDG~\cite{ldg} and the authors of cuLGT code~\cite{Schrock2013}.
The lattice QCD measurements have been performed on HOKUSAI supercomputers at RIKEN.
This work was partially supported by HPCI System Research Project
(hp120281, hp130023, hp140209, hp150223, hp150262, hp160211, hp170230, hp170170, hp180117, hp190103),
the National Key R$\&$D Program of China (Contracts Nos. 2017YFE0116700, 2018YFA0404400), the
National Natural Science Foundation of China (Grant Nos. 11935003, 11975031, 11875075, 12070131001)
, JSPS (Grant Nos. JP18H05236, JP16H03978, JP19K03879, JP18H05407), the MOST-RIKEN Joint Project ``Ab initio investigation in nuclear physics'',
``Priority Issue on Post-K computer'' (Elucidation of the Fundamental Laws and Evolution of the Universe),
``Program for Promoting Researches on the Supercomputer Fugaku'' (Simulation for basic science: from fundamental laws of particles to creation of nuclei), and Joint Institute for Computational Fundamental Science (JICFuS).

%\bibliographystyle{apsrev4-1}
%\bibliographystyle{JHEP.bst}
%\bibliography{Reference}

\begin{thebibliography}{10}

  \bibitem{Clement2017}
  H.~Clement, \emph{On the history of dibaryons and their final observation},
    \href{https://doi.org/https://doi.org/10.1016/j.ppnp.2016.12.004}{\emph{Progress
    in Particle and Nuclear Physics} {\bfseries 93} (2017) 195 }.
  
  \bibitem{Gal2015}
  A.~Gal, \emph{{Meson assisted dibaryons}},
    \href{https://doi.org/10.5506/APhysPolB.47.471}{\emph{Acta Phys. Polon. B}
    {\bfseries 47} (2016) 471}
    [\href{https://arxiv.org/abs/1511.06605}{{\ttfamily 1511.06605}}].
  
  \bibitem{Iritani2019}
  T.~Iritani, S.~Aoki, T.~Doi, F.~Etminan, S.~Gongyo, T.~Hatsuda et~al.,
    \emph{N$\omega$ dibaryon from lattice qcd near the physical point},
    \href{https://doi.org/https://doi.org/10.1016/j.physletb.2019.03.050}{\emph{Physics
    Letters B} {\bfseries 792} (2019) 284 }.
  
  \bibitem{Gongyo2018}
  {\scshape HAL QCD Collaboration} collaboration, \emph{Most strange dibaryon
    from lattice qcd},
    \href{https://doi.org/10.1103/PhysRevLett.120.212001}{\emph{Phys. Rev. Lett.}
    {\bfseries 120} (2018) 212001}.
  
  \bibitem{Huang2020}
  H.~Huang, J.~Ping, X.~Zhu and F.~Wang, \emph{{Full heavy dibaryons}},
    \href{https://arxiv.org/abs/2011.00513}{{\ttfamily 2011.00513}}.
  
  \bibitem{Richard2020}
  J.-M.~Richard, A.~Valcarce and J.~Vijande, \emph{Very heavy flavored
    dibaryons}, \href{https://doi.org/10.1103/PhysRevLett.124.212001}{\emph{Phys.
    Rev. Lett.} {\bfseries 124} (2020) 212001}.
  
  \bibitem{Junnarkar2019}
  P.~Junnarkar and N.~Mathur, \emph{Deuteronlike heavy dibaryons from lattice
    quantum chromodynamics},
    \href{https://doi.org/10.1103/PhysRevLett.123.162003}{\emph{Phys. Rev. Lett.}
    {\bfseries 123} (2019) 162003}.
  
  \bibitem{Aaij2017Omega}
  {\scshape LHCb Collaboration} collaboration, \emph{Observation of five new
    narrow ${\mathrm{\ensuremath{\Omega}}}_{c}^{0}$ states decaying to
    ${\mathrm{\ensuremath{\Xi}}}_{c}^{+}{K}^{\ensuremath{-}}$},
    \href{https://doi.org/10.1103/PhysRevLett.118.182001}{\emph{Phys. Rev. Lett.}
    {\bfseries 118} (2017) 182001}.
  
  \bibitem{Aaij2017cascade}
  {\scshape LHCb Collaboration} collaboration, \emph{Observation of the doubly
    charmed baryon ${\mathrm{\ensuremath{\Xi}}}_{cc}^{++}$},
    \href{https://doi.org/10.1103/PhysRevLett.119.112001}{\emph{Phys. Rev. Lett.}
    {\bfseries 119} (2017) 112001}.
  
  \bibitem{Bjorken1985}
  J.~Bjorken, \emph{{Is the ccc a new deal for baryon spectroscopy?}},
    \href{https://doi.org/10.1063/1.35379}{\emph{AIP Conf. Proc.} {\bfseries 132}
    (1985) 390}.
  
  \bibitem{Can2015}
  K.U.~Can, G.~Erkol, M.~Oka and T.T.~Takahashi, \emph{Look inside
    charmed-strange baryons from lattice qcd},
    \href{https://doi.org/10.1103/PhysRevD.92.114515}{\emph{Phys. Rev. D}
    {\bfseries 92} (2015) 114515}.
  
  \bibitem{Aoki2020}
  S.~Aoki and T.~Doi, \emph{Lattice qcd and baryon-baryon interactions: Hal qcd
    method}, \href{https://doi.org/10.3389/fphy.2020.00307}{\emph{Frontiers in
    Physics} {\bfseries 8} (2020) 307}.
  
  \bibitem{Ishii2007}
  N.~Ishii, S.~Aoki and T.~Hatsuda, \emph{Nuclear force from lattice qcd},
    \href{https://doi.org/10.1103/PhysRevLett.99.022001}{\emph{Phys. Rev. Lett.}
    {\bfseries 99} (2007) 022001}.
  
  \bibitem{Ishii2012}
  {\scshape HAL QCD Collaboration} collaboration, \emph{Hadron-hadron
    interactions from imaginary-time nambu-bethe-salpeter wave function on the
    lattice},
    \href{https://doi.org/https://doi.org/10.1016/j.physletb.2012.04.076}{\emph{Physics
    Letters B} {\bfseries 712} (2012) 437 }.
  
  \bibitem{Lyu2021}
  Y.~Lyu, H.~Tong, T.~Sugiura, S.~Aoki, T.~Doi, T.~Hatsuda et~al., \emph{Dibaryon
    with highest charm number near unitarity from lattice qcd},
    \href{https://doi.org/10.1103/PhysRevLett.127.072003}{\emph{Phys. Rev. Lett.}
    {\bfseries 127} (2021) 072003}.
  
  \bibitem{Ishikawa2016}
  {\scshape PACS} collaboration, \emph{{2+1 Flavor QCD Simulation on a $96^4$
    Lattice}}, \href{https://doi.org/10.22323/1.251.0075}{\emph{\textit{Proc.
    Sci.}} {\bfseries LATTICE2015} (2016) 075}
    [\href{https://arxiv.org/abs/1511.09222}{{\ttfamily 1511.09222}}].
  
  \bibitem{Aoki20013}
  S.~Aoki, Y.~Kuramashi and S.-i.~Tominaga, \emph{{Relativistic heavy quarks on
    the lattice}}, \href{https://doi.org/10.1143/PTP.109.383}{\emph{Prog. Theor.
    Phys.} {\bfseries 109} (2003) 383}
    [\href{https://arxiv.org/abs/hep-lat/0107009}{{\ttfamily hep-lat/0107009}}].
  
  \bibitem{Namekawa2017}
  {\scshape PACS} collaboration, \emph{{Charm physics by $N_f=2+1$ Iwasaki gauge
    and the six stout smeared $O(a)$-improved Wilson quark actions on a $96^4$
    lattice}}, \href{https://doi.org/10.22323/1.256.0125}{\emph{\textit{Proc.
    Sci.}} {\bfseries LATTICE2016} (2017) 125}.
  
  \bibitem{Bridge}
  \url{http://bridge.kek.jp/Lattice-code/index_e.html}.
  
  \bibitem{Doi2013}
  T.~Doi and M.G.~Endres, \emph{Unified contraction algorithm for multi-baryon
    correlators on the lattice},
    \href{https://doi.org/https://doi.org/10.1016/j.cpc.2012.09.004}{\emph{Computer
    Physics Communications} {\bfseries 184} (2013) 117 }.
  
  \bibitem{Doi2017}
  T.~Doi et~al., \emph{{Baryon interactions from lattice QCD with physical masses
    -- Overview and $S = 0, -4$ sectors --}},
    \href{https://doi.org/10.22323/1.256.0110}{\emph{\textit{Proc. Sci.}}
    {\bfseries LATTICE2016} (2017) 110}
    [\href{https://arxiv.org/abs/1702.01600}{{\ttfamily 1702.01600}}].
  
  \bibitem{Oka1987}
  M.~Oka, K.~Shimizu and K.~Yazaki, \emph{Hyperon-nucleon and hyperon-hyperon
    interaction in a quark model},
    \href{https://doi.org/https://doi.org/10.1016/0375-9474(87)90371-X}{\emph{Nuclear
    Physics A} {\bfseries 464} (1987) 700 }.
  
  \bibitem{ldg}
  \url{http://www.lqcd.org/ildg} and \url{http://www.jldg.org}.
  
  \bibitem{Schrock2013}
  M.~Schrock and H.~Vogt, \emph{Coulomb, landau and maximally abelian gauge
    fixing in lattice qcd with multi-gpus},
    \href{https://doi.org/https://doi.org/10.1016/j.cpc.2013.03.021}{\emph{Computer
    Physics Communications} {\bfseries 184} (2013) 1907 }.
  
\end{thebibliography}

%%%%%%%%%%%%%%%%%%%%%%%%%%%%%%%%%%%%%%%%%%%%%%%%%%%%%%%

%%%%%%%%%%%%%%%%%%%%%%%%%%%%%%%%%%%%%%%%%%%%%%%%%%%%%%%

\end{document}